# The Modification of the Oppenheimer and Snyder Collapsing Dust Ball to a Static Ball in Discrete Space-time

G. Chen

Donghua University, Shanghai, 201620, China
Email: gchen@dhu.edu.cn

**Abstract**. Besides the singularity problem, the famous Oppenheimer and Snyder solution is discovered to be of deficiency in two aspects: the internal Friedmann space-time does not have the inherent symmetry and cannot connect to the external Schwarzschild space-time. So the process of gravitational collapse described by this solution is doubtful. The deficiency, together with the singularity problem, result from the imperfection of the field theory in continuous space-time, which is expressed by the infinite precision function theory. The space-time structure of the Oppenheimer and Snyder dust ball is founded to be discrete rather than continuous, and to describe the field theory in discrete space-time it requires a function theory with finite precision. Based on the $i$ order real number and its equivalence class, which is defined in the real number field, the infinite precision function theory is extended to the finite precision function theory. The Einstein field equations are expressed in the form of finite precision, and then the collapsing dust ball solution in continuous space-time is modified to a static ball solution in discrete space-time. It solves all the problems of Oppenheimer and Snyder solution and shows that, with Planck length and Planck time as space-time quantum, a mechanism to resist the gravitational collapse could be obtained by the discretization of space-time.

**PACS numbers**. 02.30.-f, 04.20.Cv,04.20.Dw,04.20.Jb

## 1. Introduction

It was asserted that the gravitational collapse of dust ball results in the formation of gravitational singularity by the Oppenheimer and Snyder thesis published in 1939 [1]. On the annals of general relativity, this is a generally acknowledged milestone result. Starting with this thesis, extensive researches for the gravitational collapse and space-time singularities have been done and many theoretical hypothesis have been put forward [2-45]. However, when probing into the solution, we could find its deficiency in two aspects: the internal Friedmann space-time does not have the inherent symmetry that it should possess and cannot connect to the external Schwarzschild space-time. Therefore, the process of gravitational collapse described by this thesis is doubtful. In my opinion, the deficiencies of the Oppenheimer and Snyder solution, together with the singularity, demonstrate the imperfection of the field theory in continuous space-time, which is expressed by the infinite precision function theory. Further analysis reveals that the space-time structure of the dust ball should be discrete rather than continuous, and describing the field theory in discrete space-time requires the function theory with finite precision. So, we introduce the $i$ order real number and its equivalence class in real number field, and based on which, the infinite precision function theory is extended to the finite precision function theory. Then, we use the finite precision function theory to express the Einstein field equations and study the gravitational solution of the dust ball. By the extension and discretization of the Oppenheimer and Snyder solution, the collapsing dust ball in continuous space-time is modified to a static ball in discrete space-time. It solves all the problems of Oppenheimer and Snyder solution and shows that, with Planck length and Planck time as pace-time quantum, a mechanism to resist the gravitational collapse could be obtained by the discretization of space-time.

## 2. The problems of the Oppenheimer and Snyder solution

This section will discuss the problems of Oppenheimer and Snyder solution on the gravitational collapse of dust ball in three aspects.

Firstly, for the solution in comoving coordinates, the interior space-time geometry of dust ball can be expressed as Friedmann metric [46]

$$ds^2 = -d\bar{t}^2 + b^2(\bar{t})\left[\frac{d\bar{r}^2}{1-k\bar{r}^2} + \bar{r}^2(d\bar{\theta}^2 + \sin^2\bar{\theta}d\bar{\phi}^2)\right]. \quad (1)$$

Where $b(\bar{t})$ is given by the parametric equations of a cycloid:

$$\bar{t} = \frac{\eta + \sin\eta}{2\sqrt{k}} \quad , \quad b = \frac{1}{2}(1+\cos\eta), \quad (2)$$

For Eq.（1）, with $\eta$ defined in $[0, \pi]$, there is an inherent singularity at $\eta = \pi$. According to Birkhoff theorem, the exterior space-time geometry of the dust ball can be expressed as Schwarzschild metric [47]

$$ds^2 = -\left(1 - \frac{2GM}{r}\right)dt^2 + \frac{dr^2}{1-2GM/r} + r^2\left(d\theta^2 + \sin^2\theta d\phi^2\right). \quad (3)$$

For Eq.（3）, there are two singularities, one inherent singularity at $r = 0$, another coordinate singularity at $r = 2GM$. As we all know, the existence of the singularities reflects the incompleteness of space-time geometry in the Oppenheimer and Snyder solution.

Secondly, the interior space-time and exterior space-time of the dust ball are related to each other by the following coordinate transformations (for Eq. (1), giving $\bar{r} = a$, $k = \dfrac{2GM}{a^3}$, $a$ is the radius of dust ball in comoving coordinates, and $M$ is the gravitational mass of dust ball ) [48]:

$$r = aR(\eta) = \frac{a}{2}(1+\cos\eta), \tag{4}$$

$$t = 2GM\left\{\ln\left|\frac{(a/(2GM)-1)^{1/2}+\tan(\eta/2)}{(a/(2GM)-1)^{1/2}-\tan(\eta/2)}\right| + (a/(2GM)-1)^{1/2}\left[\eta + (a/(4GM))(\eta+\sin\eta)\right]\right\} \tag{5}$$

$$\theta = \bar{\theta}, \tag{6}$$

$$\phi = \bar{\phi}. \tag{7}$$

From the viewpoint of gravitational collapse, Eqs.(4),(5),(6) and (7) are regarded as the geodesic equations which describe the evolvement of collapsing spherical surface of the dust ball. For the given $\theta$ and $\phi$, these equations represent a radial geodesic in the $r$-$t$ plane, starting from the maximum spherical surface $r = ab(0) = a$ and terminating at the inherent singularity $r = ab(\pi) = 0$. For the geodesic, the radial coordinate $r$ is the function of coordinate time $t$, but for the Schwarzschild metric described in Eq. (3), $r$ and $t$ are two independent coordinate variables, which form a $r$-$t$ plane for the given $\theta$ and $\phi$. Of course, the geodesic in the $r$-$t$ plane is not equivalent to the $r$-$t$ plane. From Eqs. (4) and (5) we can deduce

$$\left(\frac{dr}{dt}\right)^2 = \left(1-\frac{2GM}{r}\right)^2 \frac{\left(\dfrac{r}{2GM}\right)^2\left(\dfrac{a}{r}-1\right)}{\dfrac{a}{2GM}-1} \tag{8}$$

Then substituting Eq. (8) into (3), we have

$$ds^2 = \begin{cases} -\left(1-\dfrac{2GM}{r}\right)\left(1-\dfrac{\left(\dfrac{r}{2GM}\right)^2\left(\dfrac{a}{r}-1\right)}{\dfrac{a}{2GM}-1}\right)dt^2 + r^2(d\theta^2 + \sin^2\theta d\phi^2), & r > 2GM \\[2em] -\dfrac{1}{\dfrac{2GM}{r}-1}\left(1+\dfrac{\dfrac{a}{2GM}-1}{\left(\dfrac{2GM}{r}-1\right)\left(\dfrac{r}{2GM}\right)^2\left(\dfrac{a}{r}-1\right)}\right)dr^2 + r^2(d\theta^2 + \sin^2\theta d\phi^2), & r < 2GM \end{cases} \quad (9)$$

Notice that the item $-\left(1-\dfrac{2GM}{r}\right)\left(1-\dfrac{\left(\dfrac{r}{2GM}\right)^2\left(\dfrac{a}{r}-1\right)}{\dfrac{a}{2GM}-1}\right)dt^2$ or

$-\dfrac{1}{\dfrac{2GM}{r}-1}\left(1+\dfrac{\dfrac{a}{2GM}-1}{\left(\dfrac{2GM}{r}-1\right)\left(\dfrac{r}{2GM}\right)^2\left(\dfrac{a}{r}-1\right)}\right)dr^2$ is time-like, and from Eqs. (2) and

(4) we know that the inherent time $\bar{t}$ increases when $r$ varies from $a$ to $0$, then Eq. (9) represents a keeping shrinking spherical surface rather than the Schwarzschild space-time. This means that the Schwarzschild space-time could not be formed by the process of gravitational collapse of the dust ball.

On the other hand, supposing the exterior of the dust ball is a Schwarzschlid space-time, the interior a Friedmann space-time, and the two space-time are connected by the collapsing spherical surface of the dust ball. The equation of time-like radial geodesic in the exterior Schwarzschlid space-time can be deduced as

$$\left(\frac{dr}{dt}\right)^2 = \left(1-\frac{2GM}{r}\right)^2\left(1-\frac{1}{\beta^2}\left(1-\frac{2GM}{r}\right)\right) \quad (10)$$

or

$$\left(\frac{dR_o}{dT_o}\right)^2 = \left(1-\frac{1}{\beta^2}\left(1-\frac{2GM}{r}\right)\right) \quad (11)$$

Where, $dR_o = \dfrac{dr}{\sqrt{1-\dfrac{2GM}{r}}}$, $dT_o = \sqrt{1-\dfrac{2GM}{r}}\,dt$. While the equation of time-like

radial geodesic in the interior Friedmann space-time can be deduced as

$$\left(\frac{d\bar{r}}{d\bar{t}}\right)^2 = b^{-2}(1-k\bar{r}^2)\left(1-\frac{1}{\eta^2}\right) \tag{12}$$

or

$$\left(\frac{dR_i}{dT_i}\right)^2 = \left(1-\frac{1}{\eta^2}\right) \tag{13}$$

Where, $dR_i = \frac{bd\bar{r}}{\sqrt{1-k\bar{r}^2}}$, $dT_i = d\bar{t}$. It turns out that, for Eqs. (10),(11),(12) and (13), there always exist such values of $a$, $\beta^2$, $t_a$ and $t_b$, that make the collapsing spherical surface of the dust ball locate at $r = r_a$ ($r = r_b$) when $t = t_a$ ($t = t_b$), and having $t_a < t_b$, $r_a > r_b$, $r_b > 2GM$, $\beta^2 = \eta^2\left(1-\frac{2GM}{r_a}\right)$, $\beta^2 \neq \eta^2\left(1-\frac{2GM}{r_b}\right)$. The corresponding time-like radial geodesic enters into the interior space-time from the exterior space-time at the spherical surface of $r = r_a$, and then from the interior space-time to the exterior space-time at the spherical surface of $r = r_b$. The geodesic is continuous on the spherical surface of $r = r_a$: $\left(\frac{dR_i}{dT_i}\right)_a = \left(\frac{dR_o}{dT_o}\right)_a$, and discontinuous on the spherical surface of $r = r_b$: $\left(\frac{dR_o}{dT_o}\right)_b \neq \left(\frac{dR_i}{dT_i}\right)_b$. This means that, as the collapsing of the dust ball, the continuity between the exterior and interior space-time would be broken. Therefore, in this circumstance, the Friedmann space-time and the Schwarzschild space-time could not be connected together.

Thirdly, for equation (2), when $\bar{t} = 0$, we have $b(\bar{t}) = 1$. Then, from Eq. (1), we obtain the expression of interior space geometry of the dust ball as

$$d\sigma^2 = \frac{d\bar{r}^2}{1-\kappa\bar{r}^2} + \bar{r}^2(d\bar{\theta}^2 + \sin^2\bar{\theta}d\bar{\phi}^2). \tag{14}$$

By the following demonstration, we can come to the conclusion: for $0 \leq \bar{r} \leq a$, this space geometry is part of the 3-dimension hypersphere in 4-dimensional Euclid space. At first, we make the transformation from polar coordinates to Cartesian coordinates

$$x^1 = \bar{r}\sin\bar{\theta}\cos\bar{\phi},$$

$$x^2 = \bar{r}\sin\bar{\theta}\sin\bar{\phi} \ , \tag{15}$$

$$x^3 = \bar{r}\cos\bar{\theta} \ 。$$

and let

$$x^\mu x^\mu = x^i x^i + (x^4)^2 = \frac{1}{\kappa} \ , \quad \mu = 1,2,3,4 \ , \quad i = 1,2,3 \ , \tag{16}$$

then, we can obtain the equivalent form of Eq. (14) as

$$d\sigma^2 = dx^\mu dx^\mu = dx^i dx^i + (dx^4)^2 \tag{17}$$

with

$$\bar{r}^2 = [\bar{r}(x^4)]^2 = x^i x^i = \frac{1}{\kappa} - (x^4)^2 \ 。 \tag{18}$$

We know that, if and only if $0 \leq (x^\mu)^2 \leq k^{-1}$ with $\mu = 1,2,3,4$, Eq. (16) expresses a 3-dimensional hypersphere in 4-dimension Euclid space. However, from Eq. (18), $0 \leq \bar{r} \leq a$, we can derive $k^{-1} - a^2 \leq (x^4)^2 \leq k^{-1}$, and for $a^2 < k^{-1}$, Eq. (14) is just part of the 3-dimension hypersphere.

It is well known that a space geometry with symmetry i.e., uniformity and isotropy, can be such a 3-dimension hypersphere in 4-dimensional Euclid space or a 2-dimensional sphere in 3-dimensional Euclid space, and so on. However, when $\bar{t} = 0$, the interior space geometry of the dust ball is part of the 3-dimensional hypersphere in 4-dimensional Euclid space, thus breaks the inherent symmetry.

In my opinion, the above-mentioned problems of the Oppenheimer and Snyder solution result from the imperfection of the field theory in continuous space-time, which is expressed by the infinite precision function theory. To solve these problems and obtain a complete gravitational solution of the dust ball, it requires a field theory in discrete space-time, which can be expressed by the finite precision function theory. Then, the finite precision function theory will be put forward in the next section.

## 3. Finite precision function theory

1）**The $i$ order real number and its equivalence class**

An $i$ order real number is defined as $A_i = \alpha_i C^i$. Where $C$, a real scale coefficient, is positive and greater than 1, while $\alpha_i$ a real number satisfying $C^{1/2} > |\alpha_i| \geq C^{-1/2}, i \subset \{0, \pm 1, \pm 2, \ldots\}$.

An $i'$ order equivalence class of $i$ order real number is defined as a set of $i$ order real numbers, in which the absolute value of the difference between any two $i$ order real numbers $A_i$ and $B_i$ is less than or equal to a given $i' \leq i$ order positive real number. The $i'$ order equivalence class is denoted as $\widetilde{A}_{ii'}$ or $\widetilde{B}_{ii'}$. While the relation of $A_i$ and $B_i$ is referred as $i'$ order equivalence and is denoted as $A_i \stackrel{i'}{\approx} B_i$.

Notice that, with any $i$ order real number $A_i = \alpha_i C^i$ as a reference point, different $i'$ order equivalence class $\widetilde{A}_{ii'}$ can be formed in the real number fields. An $i'$ order equivalence class of $i$ order real number can be expressed as a set of $i''$ order equivalence classes, while an $i''$ order equivalence classes can be expressed as a set of $i'''$ order equivalence classes and so on, as long as $i' \geq i'' \geq i''' \geq \cdots$. Moreover, the equivalent relation of real numbers only exists within one equivalence class that they belong to. Different equivalence classes don't have the equivalent relation.

The real number fields and the whole equivalence classes of real numbers are denoted as $R$ and $\widetilde{R}$ respectively.

**2) Function**

A function is defined as a mapping between the independent variable and dependent variable. The independent variable is called variable for short, and the dependent variable is regarded as function. The variable and function are the sets of $\widetilde{R}$ 。

**3) Operations of function**

*a*. **Limit and continuity**

**Limit**

Let $f(x)$ be an $i$ order function of $j$ order variable $x$, A an $i$ order real number, and $x_0$ a $j$ order real number. If there exists a $j'$ order equivalence class of $x_0$ and an $i'$ order equivalence class of A, when $x$ belongs to the $j'$ order equivalence class of $x_0$, $f(x)$ belongs to the $i'$ order equivalence class of A:

$$f(x) \stackrel{i'}{\approx} A , \quad x \stackrel{j'}{\approx} x_0 , \quad i' \leq i, \quad j' \leq j, \tag{19}$$

then, we define A as the limit of $f(x)$ as $x$ approaches $x_0$.

**Continuity**

Let an $i$ order function $f(x)$ be defined on a set of $j$ order real number of $\tilde{R}$, and for an element $x_0$ of the set, the value of the $i$ order function is $f(x_0)$. If for a $j'$ order equivalence class of $x_0$, there exists an $i'$ order equivalence class of $f(x_0)$, when $x$ belongs to the $j'$ order equivalence class of $x_0$, $f(x)$ belongs to the $i'$ order equivalence class of $f(x_0)$:

$$f(x) \stackrel{i'}{\approx} f(x_0) , \quad x \stackrel{j'}{\approx} x_0, \quad i' \leq i, \quad j' \leq j, \tag{20}$$

then the function $f(x)$ is continuous at $x_0$.

*b*.**Differential and derivative**

Let $f(x)$ be a $k$ order function of $j$ order variable $x$. When $x$ belongs to a $j''$ order equivalence class of a $j'$ order equivalence class, and $\Delta x$ is the difference between $x$ and any element of $j''$ order equivalence class adjacent to the $j''$ order equivalence class it belongs to, then we call

$$dy \stackrel{k''}{\approx} f(x + \Delta x) - f(x), \quad k'' \leq k' \leq k \tag{21}$$

the differential of $k$ order function $f(x)$, and

$$dx \stackrel{j''}{\approx} \Delta x, \qquad j'' \le j' \le j \qquad (22)$$

the differential of $j$ order variable. If for any $\Delta x$, there exists $i'$ ($i' \le i$) equivalence class of $i$ order real number $f'(x)$

$$f'(x) \stackrel{i'}{\approx} \frac{dy}{dx} \stackrel{i'}{\approx} \frac{f(x+\Delta x) - f(x)}{\Delta x}, \qquad (23)$$

then $f'(x)$ is called the derivative of $f(x)$.

If $f'(x)$ exists on every point of a real number set of $\tilde{R}$, then $f'(x)$ is called the derivative function or the derivative of $f(x)$.

c. **Indefinite integral and definite integral**

**Indefinite integral**

If an $i$ order function is defined in a set of $j$ order real number of $\tilde{R}$, and $f(x)$ is the derivative of a $k$ order function $F(x)$, or $f(x)dx$ the differential of $F(x)$, that is

$$f(x) \stackrel{i'}{\approx} F'(x) \stackrel{i'}{\approx} \frac{F(x+\Delta x) - F(x)}{\Delta x}, \qquad i' \le i$$

or

$$dF(x) \stackrel{k''}{\approx} f(x)dx, \qquad k'' \le k' \le k,$$

then $F(x)$ is called a primary function of $f(x)$.

It is easy to prove that, if $F(x)$ is a primary function of $f(x)$, then $F(x) + \eta$ is also a primary function of $f(x)$, where $\eta \stackrel{k'}{\approx} D$, and $D$ is any real constant.

We call all the $k$ order primary function $F(x) + \eta$ of $i$ order function $f(x)$ the indefinite integral of $f(x)$, denoted as

$$\int f(x)dx \overset{k'}{\approx} F(x)+\eta \qquad (24)$$

**Definite integral**

Let $f(x)$ be an $i$ order function of a set of $j$ order real number of $\widetilde{R}$. $a$ and $b$ are two real number of the set and $a<b$, $[a,b]$ is a subset of this set, containing all $j'$ ($j' \leq j$) order equivalence class of $j$ order real numbers, which are less than or equal to $b$ and greater than or equal to $a$. Every $j'$ order equivalence class of the subset contains a set of $j''$ ($j'' \leq j'$) order equivalence class, and the number of all the $j''$ order equivalence classes in $[a,b]$ is $\omega+1: \widetilde{A}_{jj''l} = \{\widetilde{x}_l\}, l=0,1,...,\omega$. Except the $j''$ order equivalence class containing $a$ or $b$, we take any real number $x_l \in \{\widetilde{x}_l\}$, $l=1,2,......,\omega-1$ in every $j''$ order equivalence class, and form a real number serial

$$\Delta_\omega \overset{j''}{\approx} \{x_0, x_1, x_2, ..., x_\omega\},$$
$$a \overset{j''}{\approx} x_0 < x_1 < x_2 < ... < x_\omega \overset{j''}{\approx} b.$$

Then, we define $\Delta x_l \overset{j''}{\approx} x_l - x_{l-1}$, $l=1,2,...,\omega$, and form the $k$ order sum

$$\sum_{l=1}^{\omega} f(x_{l-1})\Delta x_l \quad \text{or} \quad \sum_{l=1}^{\omega} f(x_l)\Delta x_l.$$

If for different real number serial $\Delta_\omega$, the $k$ order sums are $k'$ ($k' \leq k$) order equivalence, they are defined to be definite integral of $f(x)$ on $[a,b]$, expressed as

$$\int_a^b f(x)dx \overset{k'}{\approx} \sum_{l=1}^{\omega} f(x_{l-1})\Delta x_l \overset{k'}{\approx} \sum_{l=1}^{\omega} f(x_l)\Delta x_l. \qquad (25)$$

The above-mentioned definitions can be directly extended to the multivariable function.

## 4) The relationship between the finite precision function theory and the infinite precision function theory

It is obvious that the function theory, founded on the equivalence classes set of real numbers, should be a mathematical theory of finite precision, and the now-existing function theory, founded on the real number field, is a mathematical theory of infinite precision. The former function theory is the extension of the latter one. Firstly, a infinite precision function can always be extended to a finite precision function. For example, considering a continuous function of infinite precision, defined on the interval $[a,b]$ of real number fields, with a given scale coefficient $C$, we discretize its $i$ order variable into a set, consisting of a series of $i''$ ($i'' = i' - 1$) order equivalence classes, and the number of the equivalence classes is $\omega + 1$, two of which containing the terminals of the interval $a$ and $b$ respectively. Either of the two $i''$ order equivalence classes, combined with the adjacent $i''$ order one, constitutes a $i'$ ($i' = i - 1$) order equivalence class, and for the other $i''$ order equivalence classes in the set, the adjacent three ones constitute a $i'$ ($i' = i - 1$) order equivalence class. Accordingly, the values of the continuous function are discretized and extended to a set of $i'$ order equivalence class consisting of $i''$ order equivalence classes. Thus, we obtain a finite precision function whose variables and values are real number equivalence classes. This finite precision function has all the properties of the infinite precision one, such as limit, continuity, derivative, differential, integral, and so on. Secondly, for the variable of real number equivalence class with given order, such as the variable of $-1$ ($i' = -1$) order equivalence class of $0$ ($i = 0$) order real numbers, when $C \to \infty$, we have $\omega \to \infty$. Then the finite precision function would degenerate into the infinite precision function. On the other hand, for a given scale coefficient $C$, if $i'$ can be any integers, then the equivalent relation "$\stackrel{i'}{\approx}$" in finite precision function theory is the same as the equal relation "=" in infinite precision function theory. Based on the above-mentioned processing, we could accomplish the following conclusions: on one hand, the finite precision function theory includes the infinite precision one. On the other hand, for any differential equations of infinite precision, we can keep their forms unchanged and endow their functions, variables and operation symbols with the meaning of finite precision function theory, therefore making them become the differential equations of finite precision. Thirdly, with different scale coefficient $C$, order $i$, and equivalence classes, a finite precision function can possesses different precision or uncertainty. In some conditions, a function of finite precision or uncertainty can be degenerated into a function of infinite precision or certainty. Thus, the finite precision function theory can describe both certainty and uncertainty systems. Finally, as the range or domain of the finite precision function, the real number equivalence classes are generally the sets of discrete dots in the real number axis, which can be denoted approximately as a continuum in the special circumstance. Then, the finite precision function theory can not only denote the discrete system in the general meaning, but also describe the continuous system in the special case. So the unity of discreteness and continuity can be achieved.

## 4. A static ball in discrete space-time

On the basis of above discussion, we can extend the now-existed field theory in continuous space-time to the field theory in discrete space-time by keeping the forms of field equations unchanged, and changing a field variable or space-time coordinate from the set of $R$ to the set of $\tilde{R}$, or changing the range of value of any $n$ dimension physical field from the set of $R^n$ to the set of $\tilde{R}^n$, and changing the domain of definition of the field i.e., the 4 dimension space-time coordinates from the set of $R^4$ to the set of $\tilde{R}^4$, then changing all the operations, defined in the infinite precision continuous function theory, to those defined in the finite precision discrete function theory. In this way, the Einstein field equations of finite precision can be accomplished. Therefore, the gravitational problems of the dust ball can be solved on the basis of the Einstein field theory in discrete space-time.

For the solution of the dust ball, considering that the Schwarzschild external solution is a static metric, and the Friedmann internal solution is a dynamic metric, in order to connect the two metrics, the only way is to extend the Friedmann internal solution periodically and discretize the space-time. That is extending the domain of $\eta$ from $[0,\pi]$, defined in Eq.(2), to $R$ and discretizing it. Thus, as $\eta$ increases, the inherent time $\bar{t}$ will increases monotonously, as long as the metric factor $b$ is invariable in the meaning of equivalence, the Friedmann internal solution becomes the equivalent static one. Then, from coordinate transformations (4),(5),(6) and (7), the Friedmann internal solution can be connected to the discretized Schwarzschild external solution.

As mentioned above, by means of extension and discretization, we can endow Eqs. (1),(2),(3), (4),(5),(6) and (7) with the meaning of finite precision function theory, in which the equal sign "=" is regarded as the equivalent sign "$\stackrel{i'}{\approx}$", where $i'$ is given a proper value in every equation. Hereinafter, we will use equal sign "=" to represent the equivalent sign "$\stackrel{i'}{\approx}$" in which $i'$ can be of any integer. For the solution of dust ball with finite precision, we regard $2\pi$ as a unit element in the meaning of $-1$ $(i'=-1)$ order equivalence, choosing $\eta=0$ as a reference point, then discretizing the parameter $\eta$. That is leting $\eta = 2n\pi + \eta_{-1}C^{-1}$, $n=0,\pm1,\pm2,......,\pm N$, where $N$ an arbitrary positive integer, $C$ large enough, and for a given $n$, by changing the value of $\eta_{-1}C^{-1}$ to form an $-1$ order equivalence class of $\eta$. Considering $\overset{-1}{\eta}\approx 2n\pi$ and substituting it into the second Eq. of (2), we have $\overset{-1}{b}\approx 1$, and into Eqs. (4), (5), we have

$$\overset{-1}{r}\approx a ,\qquad(26)$$

$$\overset{-1}{t}\approx n\tau , \qquad \overset{-1}{\tau}\approx 4\pi GM(a/(2GM)-1)^{1/2}(a/(4GM)+1). \qquad(27)$$

Then, substituting $\overset{-1}{\eta}\approx 2n\pi$ and $\overset{-1}{k}\approx \dfrac{2GM}{a^3}$ into the first Eq. of (2), gives

$$\overset{-1}{\bar{t}}\approx n\bar{\tau},\qquad \overset{-1}{\bar{\tau}}\approx \pi a\sqrt{\dfrac{a}{2GM}} . \qquad(28)$$

When $\overset{-1}{\eta}\approx 2n\pi$ and $\overset{-1}{\bar{r}}\approx a$, from Eqs. (1), (2), (3), (6) and (7), we obtain

$$d\overset{-2}{\bar{t}}\approx (1-2GM/a)^{1/2}dt. \qquad(29)$$

And from Eqs. (27), (28), we can derive

$$\overset{-1}{t}\approx (1-2GM/a)^{1/2}(1+4GM/a)\bar{t} . \qquad(30)$$

Then, differentiating Eq. (30) and comparing with (29), we obtain

$$\overset{-1}{a}\approx 4GM . \qquad(31)$$

By substituting Eq.(31) into Eqs. (26),(27),and (28), we obtain

$$\overset{-1}{r}\approx 4GM ,\qquad(32)$$

$$\overset{-1}{t}\approx n\tau , \qquad \overset{-1}{\tau}\approx 8\pi GM \qquad(33)$$

and

$$\overset{-1}{\bar{t}}\approx n\bar{\tau}, \qquad \overset{-1}{\bar{\tau}}\approx 4\sqrt{2}\pi GM . \qquad(34)$$

Here, Eqs. (32), (33) denote the discretization of Schwarzschild space-time coordinates, and Eq. (34) denotes the discretization of Friedmann time coordinate.

Hereinafter, we will discuss the discretization of Friedmann space coordinates. Considering that Friedmann coordinates are the comoving coordinates of dust matter, it is obvious that the dust matter does not distribute in the points of Friedmann coordinates, otherwise for the line elements related to the dust matter, there exist $d\overset{-2}{\bar{r}}\approx 0$, $d\overset{-2}{\bar{\theta}}\approx 0$ and $d\overset{-2}{\bar{\phi}}\approx 0$. This means that the Friedmann space possesses zero measure in the meaning of $-2$ order equivalence, and would not connect to the Schwarzschild space-time which has the discrete structure denoted by Eqs. (32), (33). Thus, the dust matter must possess a kind of space-time structure without inner coordinate. In order to exhibit the structure, substituting Friedmann metric in Eq.(1) into Einstein field equations, an equation on the dust matter can be derived [51], whose finite precision form is

$$\dot{b}^2 + k \stackrel{-1}{\approx} \frac{8\pi G}{3} \rho b^2 \quad . \tag{35}$$

Then, using Eq.(31) and $k \stackrel{-1}{\approx} \frac{2GM}{a^3}$, we have $k^{-1/2} \stackrel{-1}{\approx} \sqrt{2}a \stackrel{-1}{\approx} 4\sqrt{2}GM$. Considering that $b \stackrel{-1}{\approx} 1$ and $\dot{b} \stackrel{-1}{\approx} 0$, we have

$$\rho \stackrel{-1}{\approx} \frac{3k}{8\pi G} \stackrel{-1}{\approx} \frac{M}{v_q}, \qquad v_q \stackrel{-1}{\approx} \frac{4\pi}{3} a^3. \tag{36}$$

Where $M$, $\rho$ are the mass and mass density of the dust matter respectively, and $v_q$ the volume of sphere in flat space with radius $a$. It can be inferred that the space-time structure of the dust matter is an equivalent flat ball without inner coordinates. Therefore the discrete form for the radial coordinate of Friedmann space can be derived as

$$\bar{r} \stackrel{-1}{\approx} a 。 \tag{37}$$

Thus, in the meaning of $-1$ order equivalence, the internal Friedmann space is a two-dimensional sphere in three-dimensional Euclid space. We know that the two-dimensional sphere possesses the symmetry i.e., uniformity and isotropy, and connects to the external Schwarzschild space-time on the spherical surface $r \stackrel{-1}{\approx} a$. By the analysis of radial geodesic, similar to section 2, it can be proved that the space-time geometry is continuous on the surface in the meaning of $-2$ order equivalence. Moreover, considering that gravitational fields are represented by space-time metric, and space-time metrics are denoted as the mapping of space-time coordinates, there exists no space-time metrics i.e., no gravitational fields, in the equivalent dust ball without inner coordinate. This means that the dust ball has no self-gravitational interaction, and would not produce collapse, thus forming singularities.

Finally, for Eqs.(32),(33) and (34), let

$$M \stackrel{-1}{\approx} (8\pi G)^{-1/2}, \tag{38}$$

we have

$$r \stackrel{-1}{\approx} \frac{1}{2\pi}(8\pi G)^{1/2} \qquad (39)$$

and

$$t \stackrel{-1}{\approx} n\tau, \quad \tau \stackrel{-1}{\approx} (8\pi G)^{1/2}, \qquad (40)$$

$$\bar{t} \stackrel{-1}{\approx} n\bar{\tau}, \quad \bar{\tau} \stackrel{-1}{\approx} (4\pi G)^{1/2}。 \qquad (41)$$

Where $M$, $r$, $\tau$ or $\bar{\tau}$ are the mass, radius, discrete time length of a dust ball, and they are of the orders of Planck mass, Planck length, Planck time respectively. Then, from Eqs.(32), (33) and (34), it can be deduced that, for a dust ball with the mass of $m$ times the Planck mass, the radius and discrete time length would be of $m$ times the Planck length and Planck time respectively. So if we assume the ball with mass $M$ to be an elementary particle, then many elementary particles could form into a ball with great mass. It shows that, with Planck length and Planck time as the space-time quantum, a mechanism to resist the gravitational collapse could be obtained by the discretization of space-time.

## 5. Conclusion

Besides the singularity problem, the deficiencies of the famous Oppenheimer and Snyder dust solution are discovered in two aspects: the internal Friedmann space-time does not have the inherent symmetry and cannot connect to the external Schwarzschild space-time. The problems of the solution result from the imperfection of the field theory in continuous space-time, which is expressed by the infinite precision function theory. To solve these problems and obtain a complete gravitational solution of the dust ball, it requires a field theory in discrete space-time, which can be expressed by the finite precision function theory. Based on the $i$ order real number and its equivalence class, which is defined in the real number field, the infinite precision function theory is extended to the finite precision function theory. The Einstein field equations are expressed in the form of finite precision, and the Oppenheimer and Snyder collapsing dust ball solution is modified to a static ball solution. This solution shows that: a) The space-time geometry of the ball has a discrete structure. b) The matter of the ball is existed in a equivalent flat ball without inner coordinates. Thus it would not give rise to collapse and form singularities by the self gravitational interaction. c) There exists a ball of Planck mass, with Planck length and Planck time as its radius and elementary time length respectively, and the ball can be regarded as a elementary particle. d) m particles can form into a large ball, which has mass of m times the Planck mass, radius of m times the Planck length, and discrete time length of m times the Planck time. As a result, with Planck length and Planck time as pace-time quantum, a mechanism to resist the gravitational collapse could be obtained by the discretization of space-time. From above discussion, we come to the conclusion that the finite precision function theory should be adopted to accomplish a complete description of the gravitation phenomenon. Taking advantage of this mathematical theory, we can not only solve the singularity problems of classical general relativity theory, but also reveal more deep natures of the physical world.